# Structural Disorder, Octahedral Coordination, and 2-Dimensional Ferromagnetism in Anhydrous Alums

D.V. West[1], Q. Huang[2], H.W. Zandbergen[3], T.M. McQueen[1], and R.J. Cava[1]

[1]Department of Chemistry, Princeton University, Princeton NJ

[2]Center for Neutron Research, The National Institute for Standards and Technology, Gaithersberg, MD

[3]Kavli Institute for Nanoscience, Technical University of Delft, the Netherlands

## Abstract

The crystal structures of the triangular lattice, layered anhydrous alums $KCr(SO_4)_2$, $RbCr(SO_4)_2$ and $KAl(SO_4)_2$ are characterized by X-ray and neutron powder diffraction at temperatures between 1.4 and 773 K. The compounds all crystallize in the space group $P\bar{3}$, with octahedral coordination of the trivalent cations. In all cases, small amounts of disorder in the stacking of the triangular layers of corner sharing $MO_6$ octahedra and $SO_4$ tetrahedra is seen, with the $MO_6$-$SO_4$ network rotated in opposite directions between layers. The electron diffraction study of $KCr(SO_4)_2$ supports this model, which on average can be taken to imply trigonal prismatic coordination for the $M^{3+}$ ions; as was previously reported for the prototype anhydrous alum $KAl(SO_4)_2$. The temperature dependent magnetic susceptibilities for $ACr(SO_4)_2$ (A = K,Rb,Cs) indicate the presence of predominantly ferromagnetic interactions. Low temperature powder neutron diffraction reveals that the magnetic ordering is ferromagnetic in-plane, with antiferromagnetic ordering between planes below 3 K.

**Introduction**

The anhydrous alums, with generic formula $A^{1+}M(XO_4)_2$, are a class of layered materials in which tetrahedral polyatomic anions (e.g. $SO_4^{2-}$, $MoO_4^{2-}$, $PO_4^{3-}$) share corners with $MO_6$ polyhedra in layers that are separated by the larger $A^+$ ions (e.g. $K^+$, $Rb^+$, $NH_4^+$). The metal M can be a large variety of *d*, or *p*-block elements[1-9]. The $MO_6$ polyhedra, usually octahedra, are arranged such that the M ions form planes of edge-sharing triangles. Anhydrous alums crystallize in monoclinic[1], trigonal[10] and rhombohedral[2] symmetry. In the monoclinic compounds, the M-triangles are isosceles, while in the trigonal and rhombohedral compounds the triangles are equilateral, and either eclipsed or staggered, respectively, in adjacent layers.

These compounds have been proposed as ideal systems for studying magnetism on a 2D triangular lattice[11]. With relatively weak magnetic interactions due to their long M-O-X-O-M superexchange pathways, they allow for the experimental observation of the effects of frustration under weak, readily available magnetic fields (< 10 T). Some trigonal Fe-based compounds, for example, have been proposed as ideal examples of the antiferromagnetic 2D Heisenberg model on a triangular lattice[12], while monoclinic $KFe(SO_4)_2$ has an incommensurate spiral magnetic structure[13]. From the viewpoint of magnetic frustration, $Cr^{3+}$-based compounds are of particular interest due to the isotropic spin state of the $d^3$ spin 3/2 ion with octahedral coordination. This motivates the present study of $KCr(SO_4)_2$, $RbCr(SO_4)_2$ and $CsCr(SO_4)_2$.

The most significant difference in the reported crystal structures of the anhydrous alums is the nature of the $MO_6$ coordination polyhedra. The $MO_6$ coordination is octahedral in the crystal structures reported in centrosymmetric space groups[1-3,10], while it is apparently triangular prismatic for crystal structures reported in two chiral space groups[4,14]. The faces of the $MO_6$ polyhedra are parallel to the triangular layers for all structures. They only differ via a rotation of the three oxygens of the upper triangular face of the $MO_6$ polyhedron with respect to the three oxygens in the lower triangular face of the polyhedron, around an axis through the center of the polyhedron perpendicular to the triangular plane. The attached $XO_4$ tetrahedra also rotate. This distinction is shown in Figure 1. This arrangement, which appears to be trigonal prismatic coordination for the M ions, is contrary to what is expected for transition elements and other elements where octahedral $MO_6$ coordination polyhedra are favored. Because there are no steric restrictions in the anhydrous alum structure type to suggest that the ideal preferred coordination polyhedra cannot easily be attained, such structures, even that of anhydrous alum $KAl(SO_4)_2$ itself[4], must be carefully considered.

Thus, in addition to reporting the crystal structures of the previously uncharacterized Cr-based anhydrous alums $A^{1+}Cr(SO_4)_2$, ($A^{1+}$ = K, Rb, Cs), we also re-examine the crystal structure of the anhydrous alum $KAl(SO_4)_2$, previously reported in the space group P321 (with the $Al^{3+}$ in trigonal prismatic coordination with oxygen). We find that structural disorder, due to the disorder in the rotation of $MO_6$ octahedra from one layer to the next, is present in all compounds. This disorder mimics trigonal prismatic coordination, resulting in the incorrect previous structure determination for $KAl(SO_4)_2$. The Cr compounds are also characterized by measurements of the magnetizations between 1.8 and 200 K. Magnetic ordering is observed at 3.5, 2.5, and 1.8 K for the K,

Rb and Cs versions, respectively. The long-range ordered magnetic structure of KCr(SO$_4$)$_2$, determined by powder neutron diffraction at 1.4 K, is reported.

**Experimental**

The Cr$_2$(SO$_4$)$_3$•XH$_2$O (Alfa) and Al$_2$(SO$_4$)$_3$•XH$_2$O (Alfa) hydrates were dried in air at 450 °C, and the resultant anhydrous sulfates were weighed out while hot to prevent re-hydration. All samples were made with stoichiometric quantities of K$_2$SO$_4$, Rb$_2$SO$_4$, Cs$_2$SO$_4$, Cr$_2$(SO$_4$)$_3$ and Al$_2$(SO$_4$)$_3$. The powders were intimately mixed, and placed in open quartz tubes. The powder was further dried by placing the quartz tubes in a furnace at 420 °C under flowing N$_2$ gas for at least two hours. The tubes were then evacuated, sealed and placed in a furnace at 700 °C overnight, and air-quenched. Two other samples of KCr(SO$_4$)$_2$ were prepared, one by water quenching the tube from 700 °C, the other by slow cooling from 700 °C, followed by annealing at 300 °C for 10 days. The sintered samples were lightly broken down into loose powders in a mortar and pestle. Grinding after synthesis was kept to a minimum, as the samples were found to be very sensitive to stress, as described below.

Magnetic data were measured for ACr(SO$_4$)$_2$ using a Quantum Design PPMS magnetometer. Temperature dependent magnetization data were measured under a field of $\mu_0 H = 0.02$ T in the temperature range 1.8 – 20 K, and also $\mu_0 H = 1$ T in the range 1.8 – 200 K; p$_{eff}$ and $\theta_{CW}$ values were determined from fits to the Curie Weiss law in the temperature range 50-200 K. Field dependent magnetization curves were also measured up to $\mu_0 H = 9$ T at 1.8 K.

Nuclear and magnetic structures were characterized using X–ray and neutron diffraction of polycrystalline samples. X-ray powder diffraction (PXRD) data were collected on all samples (both un-ground and well ground) at room temperature with a Bruker D8 Focus, using Cu Kα radiation and a graphite diffracted beam monochromator. Neutron Powder Diffraction (NPD) data were collected on un-ground samples of $KAl(SO_4)_2$ (298 K), $RbCr(SO_4)_2$ (1.4, 8 and 298 K), and all three samples of $KCr(SO_4)_2$ (298 K for all three, 1.4 and 8 K for the air-cooled and 10-day annealed samples, and finally 773 K in an evacuated furnace for the 10-day annealed sample) at the NIST Center for Neutron Research on the high resolution powder neutron diffractometer (BT–1) with neutrons of wavelength 1.5403 Å produced by using a Cu(311) monochromator. Collimators with horizontal divergences of 15′, 20′ and 7′ of arc were used before and after the monochromator and after the sample, respectively. Data were collected in the 2θ range of 3–168° with a step size of 0.05°. The structural parameters were determined by Rietveld refinement of the neutron diffraction data using the GSAS program [15,16]. The atomic neutron scattering factors used in the refinements were K: 0.367, Rb: 0.708, Al: 0.345, Cr: 0.363, S: 0.285, and O: 0.581 x $10^{-12}$ cm.

Both un-ground and well ground polycrystalline samples of $KCr(SO_4)_2$ were studied by synchrotron X-ray diffraction. High resolution X-ray powder diffraction data were collected on the SUNY X16C beamline at the National Synchrotron Light Source. The direct synchrotron beam was monochromated by a double Si(111) crystal at a wavelength of 0.69899 Å, as determined from seven well defined reflections of an $Al_2O_3$ NIST standard. Samples were packed in 1 mm OD (0.01 mm wall) glass capillaries and measured in transmission geometry with a Ge analyzer crystal and a NaI detector. The electron diffraction study of $KCr(SO_4)_2$ was performed with a Phillips CM 200 electron microscope equipped with a field emission gun.

**Crystal Structures**

The laboratory PXRD patterns for all samples made under the conditions described displayed peak positions and Miller indices consistent with trigonal symmetry unit cells. There are three reported trigonal space groups for the anhydrous alums, distinguished only by the orientations of the XO$_4$ tetrahedra. The first is $P\bar{3}m1$, in which the MO$_6$ coordination is octahedral and the XO$_4$ tetrahedra are orientated in high symmetry directions such that their triangular bases in the crystallographic *a-b* plane are parallel to the 100, 010, or 110 directions (Figure 1a). The second is $P\bar{3}$, in which the triangular bases of the tetrahedra rotate around an axis parallel to *c*, away from the high symmetry orientations, while still maintaining the MO$_6$ octahedral coordination (Figure 1b). The last is $P321$, the only chiral trigonal space group reported, and the one reported for the prototype anhydrous alum KAl(SO$_4$)$_2$. In this space group, the triangular bases of the tetrahedra above the triangular M cation plane are rotated away from the ideal orientations, but in an opposite direction to those of the tetrahedra below the plane, resulting in distorted trigonal prismatic MO$_6$ coordination (Figure 1c).

Quantitative structure refinement based on the laboratory PXRD patterns was not possible due to preferred orientation and unusually severe effects of grinding. Forceful grinding of the Cr-based samples causes a dramatic color change in the powders from an intense green, to a very faint whitish green, and a strong increase in the 00l reflections relative to the other peaks. The NPD studies were not subject to these effects due to the sample geometry and the minimal post-synthesis grinding employed to prepare the samples. All of the diffraction patterns at all temperatures (1.4, 8, 298 and 773 K) displayed the same characteristics. As such, detailed discussion is presented below for the refinement of only one of the NPD patterns, the 298 K pattern for the water quenched $KCr(SO_4)_2$ sample, with the considerations being applicable to all of the other NPD patterns.

A test of the highest symmetry structural model for $KCr(SO4)_2$ yielded a very poor fit to the NPD data, with predicted NPD intensities not even qualitatively matching the observed diffraction pattern. The predicted neutron diffraction patterns for anhydrous alum crystal structures in the other two space groups $P321$ and $P\bar{3}$ are starkly different. Refinement of the structure of $KCr(SO_4)_2$ in the space group $P321$ also gave very poor agreement between the model and the observed diffraction data; the agreement never yielded a $\chi^2$ less than 28 (the best refinement, a different space group with a different structural model, had only one additional parameter and $\chi^2 = 1.59$), indicating that this structure is also not a possibility for this compound. While providing a much better fit, a refinement of a simple model in the space group $P\bar{3}$ was also not fully satisfactory. Aside from incorrect intensity predictions, small peaks are observed in the NPD patterns corresponding to a doubled c-axis, and a broad, weak hump-like feature is seen spanning the 2θ range 37° to 47°. The hump appears to be comprised of extra small peaks, or diffuse scattering, or some combination of both, and is modeled here with extra background parameters. Difference maps revealed the presence of extra scattering at positions in the unit cell consistent with rotations of the triangular bases of some fraction of the $SO_4$ tetrahedra (typically near 10 % in all samples). The bases of these minority tetrahedra are rotated away from the high symmetry direction, around the same axis through the apex of the tetrahedra, by an equivalent angle, but in the opposite sense, to those of the main group of tetrahedra.

These observations led us to perform an electron diffraction study of $KCr(SO_4)_2$ The electron diffraction patterns were found to be consistent with trigonal symmetry, and the hk0 diffraction plane shows the expected hexagonal diffraction pattern (Figure 2a). However, a variety of supercell reflections were observed along the *c*-axis (Figure 2b-d). The patterns showed reflections where the c-axis is doubled, tripled, and sometimes streaked, indicating different stacking sequences and disorder in stacking along c are all present in the samples.

A model including disorder in the rotational position of some of the sulfate tetrahedra yielded an excellent final fit to the diffraction data for all samples at all temperatures. The three oxygens of the triangular base of the sulfate tetrahedron are in a single 6g site in space group $P\bar{3}$ in the ideal ordered structure. To model the tetrahedra rotated in the opposite direction, a second set of tetrahedral base oxygens was added to the refinements in a different 6g site. The sum of the occupancies for these two 6g sites was constrained to unity, and without any further constraints, the spatial coordinates of the sites freely refined to have the relationship: 6g1 = (x,y,z), and 6g2 = (-y,-x,z) (Figure 3). The coordinates were for example, 6g1 = (0.0619, 0.7054, 0.3531) and 6g2 = (0.9381, 0.2945, 0.3531) for $KCr(SO_4)_2$ at room temperature, with fractional occupancies of 0.92 and 0.08 for the two types of sites, respectively. With this disorder, the agreement between the model and the observed NPD data was improved dramatically, with $\chi^2$ dropping from 3.75 to 2.15. The fit was further improved by modeling the disordered 6g oxygens with anisotropic thermal parameters. All other sites were modeled with isotropic thermal parameters, as the fit quality was unchanged by the use of anisotropic thermal parameters. The model indicates that there is disorder in the stacking of the layers; rotations of the tetrahedra within a single plane are coupled through the connections to the bases of the $MO_6$ octahedra and cannot happen in opposite sense within a single plane.

This addition correctly modeled the large majority of the sample in which there is 10% disorder in the rotation of the $SO_4$ tetrahedra on translating from one plane to the next. However both the NPD and electron diffraction patterns indicated that a number of crystallites have an ordered arrangement of the rotations along the c-axis. The doubled c-axis supercell structure is modeled by including a minor fraction of a 2nd phase that has an ordering of the layer stacking, as discussed above (Figure 4). The unit cell has two layers. The bottom layer is the simple unit cell with only the 6g1 oxygens for the sulfate groups, while the top layer is the original unit cell with only the 6g2 oxygens for the sulfates. In alternating planes, these are found in a 1:1 ratio in the ordered phase (Figure 4 right). This ordered supercell has the space group $P321$. All of the average atomic position coordinates, and the unit cell parameters of the two phases are identical or related, and were constrained such that when compared to a disordered $P\bar{3}$ model, only one additional parameter, the phase fraction of the doubled unit cell phase, is needed. These two phase refinements give excellent agreement to the NPD diffraction patterns for all of the compounds studied (Figure 5). The final structural models for all phases are presented in Table I.

Synchrotron X-ray powder diffraction was used to confirm the symmetry and stacking disordered $P\bar{3}$ model for $KCr(SO_4)_2$, and to study the effects of grinding on the sample. The pattern of the un-ground sample gave very sharp peaks, with only a few very small impurity peaks that were not indexable by the anhydrous alum structure type. There is no indication of lower symmetry in the high angle diffraction peaks (i.e. the trigonal peaks were not split). The synchrotron X-ray pattern was refined with the three different structural models as described above, in space groups $P321$, $P\bar{3}$, and $P\bar{3}$ with oxygen disorder, which gave agreement factors, $R(F^2)$, of 11.02 %, 5.22 % and 4.21 % respectively. The 6g1 and 6g2 oxygen occupancies were found to be 0.927 and 0.073 for the disordered model, within error of the fractions determined by NPD. The pattern of a $KCr(SO_4)_2$ sample that had been forcefully ground for two minutes showed severely broadened peaks with a 10-fold reduction in peak height, and a new set of peaks, the origin of which is at present unknown (Figure 6b), confirming the highly sensitive character of the structure to stresses on the order of the few kilobar applied during grinding.

A brief consideration of the inter-layer bonding provides an understanding of the ease of stacking disorder in the anhydrous alums. The 12-fold coordination of the Alkali metals by oxygens, 1 apical and 1 triangular plane oxygen from 6 different $SO_4$ groups, serves as the sole bonding between layers. Disorder in stacking will not be energetically expensive for these structures, then, if the $AO_{12}$ coordination polyhedra are not significantly different in the two stacking possibilities, i.e. when there are identical sulfate planes above and below, or rotated planes above and below the A cation layer. The primary energetic distinction between the two kinds of layer stacking comes from the relative positions of the triangular plane (6g site) oxygens in the $SO_4$ groups above and below alkali cation sheets (Figure 7). The case of identical layer stacking results in a staggered array for some of the oxygens, while alternate layer stacking has them nearly eclipsed. The A-O distances are equal in both cases and the relative O-O distances of 5.931 Å and 5.745 Å are close enough that a small amount of disorder is chemically reasonable.

The structural refinements for all compounds, including the prototype anhydrous alum $KAl(SO_4)_2$ itself, show conclusively that these compounds exhibit octahedral coordination of oxygen around the trivalent cations. The data further shows that there is typically 10% disorder in the stacking of planes of $SO_4$-$MO_6$ polyhedra due to rotations of the polyhedra. This disorder leaves the in-plane M-M distances of the magnetic ions unaffected, and further, since the geometry of the planes is identical no matter what the sense of the rotation, leaves the in-plane superexchange pathways identical, and changes only the O6g-K-O6g bond-angle in the plane-to-plane superexchange pathways. Thus the structural disorder will have essentially no impact on the magnetic interactions in compounds with the anhydrous alum structure type.

**Magnetic Characterization**

The magnetic characterizations of $KCr(SO_4)_2$, $RbCr(SO_4)_2$, and $CsCr(SO_4)_2$ reflect the 2-dimensional nature of these compounds. The temperature dependent magnetic susceptibilities between 1.8 and 200 K indicate the presence of dominantly ferromagnetic interactions for all compounds with low $\theta_{cw}$ values ~ 4 K or less. Fits to the Curie Weiss law for $1/\chi(T)$ data in the temperature range 50-200 K (Figure 8) yielded $p_{eff}$ values consistent with $S = 3/2$ and weakening interaction strength with increasing A cation size. None of the compounds shows the kind of sharp feature in the measured susceptibilities usually indicative of a long range magnetic ordering transition.

However, low temperature susceptibility data under a 0.0200 T field show plateaus typical of those seen in ferromagnets (Figure 9 inset). The plateaus shift to lower temperatures as the Cr-Cr distance increases, indicating a weakening of the ferromagnetic interactions with increasing triangular plane size on going from K to Rb to Cs, along with any accompanying changes in the Cr-O-S bond angle along the Cr-O-S-O-Cr superexchange pathway (without a NPD structure refinement of $CsCr(SO_4)_2$, a precise comparison of bond angles cannot be made). The field-dependent magnetization data at 1.8 K show clear saturations for all compounds at very low fields (less than 3 T) at the expected value of 3 $\mu B/Cr^{3+}$ (Figure 9). No magnetic hysteresis was observed in any of the samples at 1.8 K.

The low temperature NPD study of $KCr(SO_4)_2$ revealed the presence of strong magnetic peaks below 2.75 K, indexable by a unit cell with a doubled *c*-axis (Figure 10). If the ordering in these phases was purely ferromagnetic in character, then the magnetic unit cell would remain unchanged; thus the diffraction data are incompatible with simple ferromagnetic ordering. In this doubled unit cell, there are only two independent magnetic atoms, one per layer. Thus, ferromagnetic in-plane ordering, with moments either in-plane or canted from the plane, antiferromagnetically ordered from one layer to the next, are the only possibilities for the magnetic structure. Rietveld refinement of the Cr moments, constrained only to be antiparallel, yielded magnetic moments canted 27° (±2) from the *a-b* plane. However, a model in which the moments are further constrained to be exactly in the plane yielded a very similar fit, and therefore the two models cannot be distinguished. Because the symmetry is trigonal, all directions within the *a-b* plane for a magnetic moment are equivalent[17], and the depiction of the moments parallel to the 100 direction in Figure 10 is arbitrary. The intensity of the (001) reflection of the doubled magnetic unit cell was measured as a function of temperature to determine the ordering parameter, and shows an ordering temperature of 2.75 K (Figure 10, inset). The intensity of the magnetic reflections has not yet saturated at the lowest temperature of the measurements, 1.4 K, indicating that the magnetic ordering is continuing to develop at this temperature.

The clear implication from the positive $\theta_{cw}$ values and ferromagnetic-like saturation of the moments is that the dominating magnetic interactions are the in-plane ferromagnetic interactions. While the NPD data show that the layers are coupled antiferromagnetically, the field dependent magnetization data show that even in a small magnetic field, the antiferromagnetically coupled layers flip to align ferromagnetically, indicating that the inter-layer coupling is extremely weak.

**Conclusions**

The anhydrous alums have proven to be two dimensional in almost every aspect. Powder diffraction of neutrons, electrons and synchrotron X-rays are all consistent with octahedral coordination of the trivalent cations in the space group $P\bar{3}$ and rotational disorder along the c-axis. The nature of this disorder derives from the similar energies of the different rotations of the sulfate groups within a layer with respect to its nearest neighbors. The magnetic behavior is dominated by in-layer ferromagnetic interactions, which weaken with increasing A cation size, and consequently lattice size, with low temperature powder neutron diffraction indicating even weaker antiferromagnetic inter-layer coupling. Unlike the case for most transition metal oxides with triangular lattices, the magnetic interactions in the Cr-based anhydrous alums are weak enough to allow study at relatively low fields and experimentally accessible temperatures, making these good systems for studying two dimensional ferromagnets on a triangular lattice. Tuning the in-plane interactions to antiferromagnetic, if possible, should yield interesting compounds for study of geometric frustration. More detailed neutron diffraction characterization of the magnetic states of this family of compounds may also be of future interest.

**Acknowledgements**

This research was supported by the NSF program in Solid State Chemistry, grant number NSF DMR-0703095. Certain commercial materials and equipment are identified in this report to describe the subject adequately. Such identification does not imply recommendation or endorsement by the NIST, nor does it imply that the materials and equipment identified are necessarily the best available for the purpose. We gratefully acknowledge Peter W. Stephens for his assistance in obtaining the synchrotron X-ray patterns. Use of the National Synchrotron Light Source, Brookhaven National Laboratory, was supported by the U.S. Department of Energy, Office of Science, Office of Basic Energy Sciences, under Contract No. DE-AC02-98CH10886. T. M. McQueen gratefully acknowledges support of the National Science Foundation Graduate Research Fellowship Program.

**Figure Captions:**

**Figure 1. Layer comparison between reported space groups for trigonal anhydrous alums. a)** Space Group $P\bar{3}m1$ has the polyhedra in high symmetry orientations such that the bases of the sulfate tetrahedra are aligned with the 100, 010 and 110 directions.  **b)** $P\bar{3}$ has the sulfates above and below the Cr layer rotated about an axis parallel to 001 equally and in the same direction.  **c)** $P321$ has the sulfates rotated as in $P\bar{3}$, except that the sulfates above the $M^{3+}$ plane are rotated equally in degree but in an opposite direction to the those below the $M^{3+}$ plane, yielding trigonal prismatic coordination of the trivalent cations.

**Figure 2. TEM diffraction images of $KCr(SO_4)_2$.  a)** hk0 plane showing the expected hexagonal diffraction pattern.  Diffraction images of different parts of a single crystallite showing doubling (**b**), tripling (**c**) and some streaking (**d**) along the $c$-axis.

**Figure 3. Model of $KCr(SO_4)_2$ showing 6g oxygen disorder.** A second 6g oxygen sits across a 110 pseudo mirror plane as a result of sulfate rotational disorder between layers. The O2 and O3 occupancies are approximately 90%:10% respectively.

**Figure 4. Stacking faults in anhydrous alums.**  The in-plane polyhedra orientations are held in place by a rigid bonding network meaning that 6g oxygen disorder is a manifestation of stacking faults between layers, not disorder within the layers.

**Figure 5. Neutron powder diffraction of anhydrous alums.**  Refinements of two phase models for all diffraction patterns gave excellent fits with consistent values across the series for 6g oxygen disorder and phase percentages of the doubled $c$-axis phase. Asterisks mark the doubled $c$-axis peaks resulting from stacking faults.

**Figure 6. Synchrotron X-ray powder diffraction of un-ground and well ground**

**KCr(SO$_4$)$_2$.** Refinement of the un-ground sample in $P\bar{3}$ shows 6g oxygen disorder consistent with the NPD patterns. The diffraction pattern of the well ground sample shows a 10-fold drop in peak height, due to broadening of the peaks, and the appearance of another phase.

**Figure 7. Comparison of interlayer bonding between anhydrous alums with and without stacking faults.** The energy cost in a stacking fault comes from the eclipsing of the 6g oxygens on either side of the A cation layer. The energies of the stacking sequences are comparable because of the large separation between the 6g oxygens. The rigid in-layer bonding prevents the sulfates from rotating back to their ideal positions.

**Figure 8. 1/χ data (μ$_0$H = 1 T) for Cr-based alums.** Fits to the Curie Weiss law were done from 50-200 K. All compounds are consistent with spin 3/2 and show ferromagnetic θ$_{cw}$ values which decrease with increasing A cation radius.

**Figure 9. Low temperature and low field magnetization data.** M v. H data at 1.8 K show that all compounds saturate at the expected 3 μB per Cr$^{3+}$. χ v. T data (inset) under a 0.02 T field show the decreasing ferromagnetic interactions as a function of A cation radius.

**Figure 10. Magnetic characterization by low temperature neutron powder diffraction. a)** NPD structure refinement at 1.4 K. **b)** The bulk of the magnetic intensity occurs at 2θ = 5.5, the 001 reflection for a doubled c-axis cell indicating antiferromagnetic ordering between the layers. **c)** A plot of the 001 magnetic intensity v. T shows that antiferromagnetic ordering, which onsets at 2.75 K, is still developing at 1.4 K. The line is drawn to guide the eye. **d)** Diagram showing the ferromagnetically ordered layers aligned antiferromagnetically with respect to eachother.

**Table I. Refined Structural Parameters for Anhydrous Alums**

**Simple Cell Parameters**. Space Group: $P\bar{3}$. $A^{1+}$ in 1a (0,0,0). $M^{3+}$ in 1b (0,0,$^1/_2$). S 2d ($^1/_3$,$^2/_3$,z). O1 2d ($^1/_3$,$^2/_3$,z). O2 6g (x,y,z). O3 6g, related to O2 by (-y,-x,z).

**Doubled Cell Paramters.** Space Group: $P321$. $A^{1+}1$ 1a (0,0,0). $A^{1+}2$ 1b (0,0,$^1/_2$). $M^{3+}$ 2c (0,0,$^1/_4$). S1 2d, related to Simple Cell (SC) S by ($^1/_3$,$^2/_3$,$^z/_2$). S2 2d, related to SC S by ($^2/_3$,$^1/_3$,$^{(-z)}/_2$). O1 6g, related to SC O1 by ($^1/_3$,$^2/_3$,$^z/_2$). O2 6g related to SC O1 by ($^2/_3$,$^1/_3$,$^{(-z)}/_2$). O3 6g, related to SC O2 by (x,y,$^z/_2$). O4 6g, related to SC O2 by (-x,-y,$^{(-z)}/_2$)

| atom | site symm | param. | KAl(SO$_4$)$_2$ neutron 298 K | KCr(SO$_4$)$_2$ neutron 298 K | KCr(SO$_4$)$_2$ synchrotron 298 K | KCr(SO$_4$)$_2$ neutron 1.4 K | RbCr(SO$_4$)$_2$ neutron 298 K |
|---|---|---|---|---|---|---|---|
|  |  | a | 4.7206(1) | 4.7561(1) | 4.7624(1) | 4.7463(1) | 4.7850(1) |
|  |  | c | 7.9835(1) | 8.0483(1) | 8.0578(1) | 7.9708(1) | 8.3685(1) |
| $A^{1+}$ | 1a | $B_{iso}$ | 1.78(6) | 1.77(6) | 1.33(4) | 0.42(5) | 1.82(2) |
| $M^{3+}$ | 1b | $B_{iso}$ | 0.75(4) | 0.75(4) | 0.28(2) | 0.23(4) | 0.88(3) |
| S | 2d | z | 0.2964(3) | 0.2907(3) | 0.2911(2) | 0.2896(1) | 0.2999(3) |
|  |  | $B_{iso}$ | 0.87(4) | 0.81(4) | 0.39(3) | 0.15(4) | 0.91(3) |
| O1 | 2d | z | 0.1169(2) | 0.1119(2) | 0.1116(5) | 0.1072(2) | 0.1274(2) |
|  |  | $B_{iso}$ | 1.48(2) | 1.56(2) | 1.11(8) | 0.43(2) | 1.48(2) |
| O2 | 6g | x | 0.0698(2) | 0.0619(2) | 0.0647(5) | 0.0593(1) | 0.0650(1) |
|  |  | y | 0.7206(2) | 0.7054(2) | 0.7099(5) | 0.7036(2) | 0.7088(2) |
|  |  | z | 0.3599(1) | 0.3531(1) | 0.3527(3) | 0.3513(1) | 0.3598(1) |
|  |  | $B_{11}$ | 0.73(2) | 0.78(2) | $B_{iso}$ - | 0.15(3) | 0.94(2) |
|  |  | $B_{22}$ | 0.81(2) | 0.92(3) | 0.50(6) | 0.20(2) | 0.84(2) |
|  |  | $B_{33}$ | 1.77(2) | 1.89(3) | - | 0.50(2) | 1.75(2) |
|  |  | Occ. | 0.930(2) | 0.918(2) | 0.927(3) | 0.920(2) | 0.936(1) |
| O3 | 6g | Occ. | 0.070(2) | 0.082(2) | 0.073(3) | 0.080(2) | 0.064(1) |
|  | Doubled cell phase % |  | 1.61% | 2.20% | - | 1.70% | 1.99% |
|  |  | $\chi^2$ | 1.75 | 1.59 | 3.03 | 3.35 | 2.074 |
|  |  | Rp | 4.63% | 4.21% | 12.25% | 5.33% | 3.94% |
|  |  | Rwp | 5.88% | 5.29% | 8.80% | 6.86% | 4.91% |
|  |  | R(F$^2$) | 3.60% | 3.87% | 4.21% | 3.55% | 3.68% |

**Figure 1.**

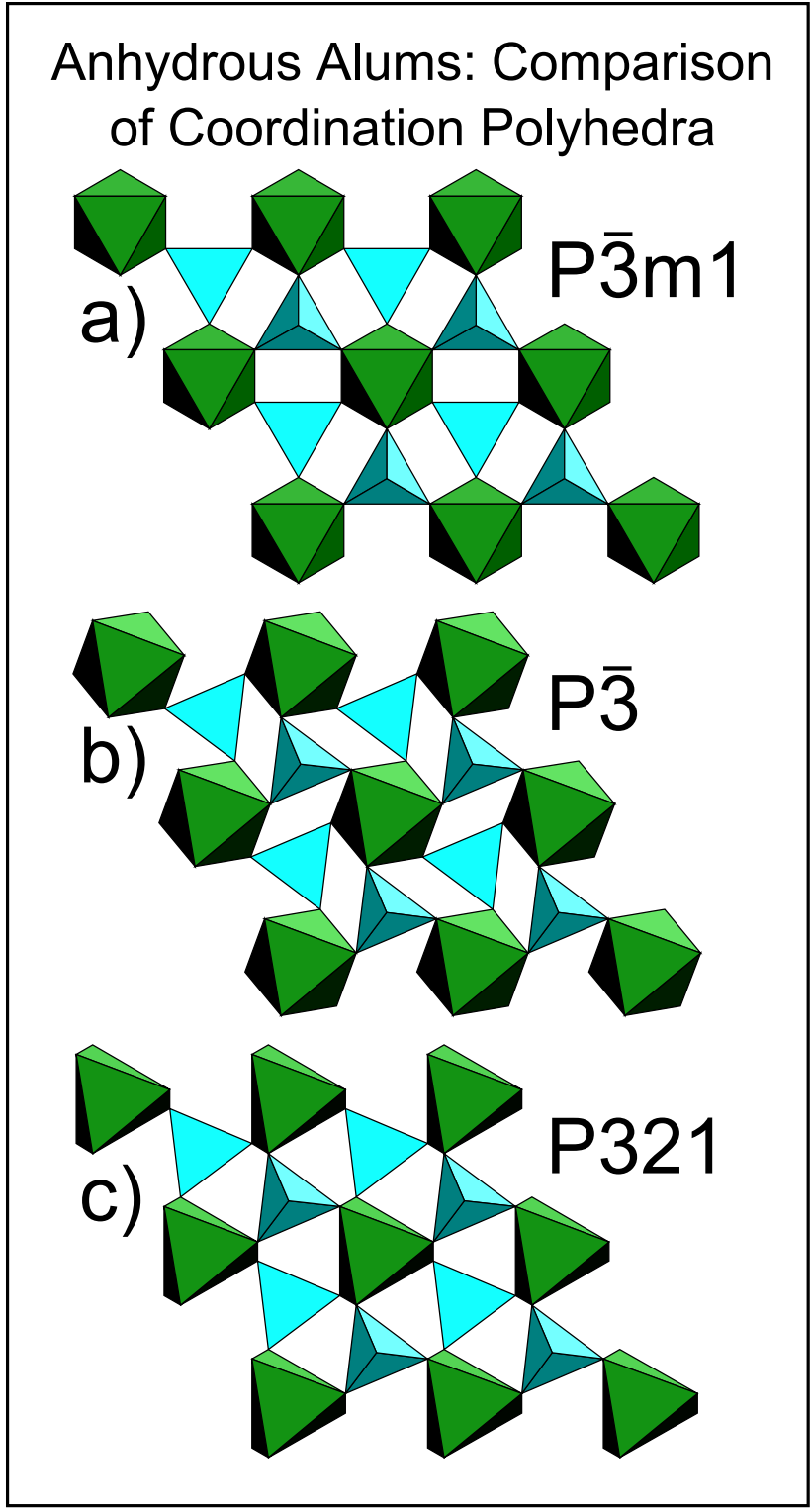

**Figure 2.**

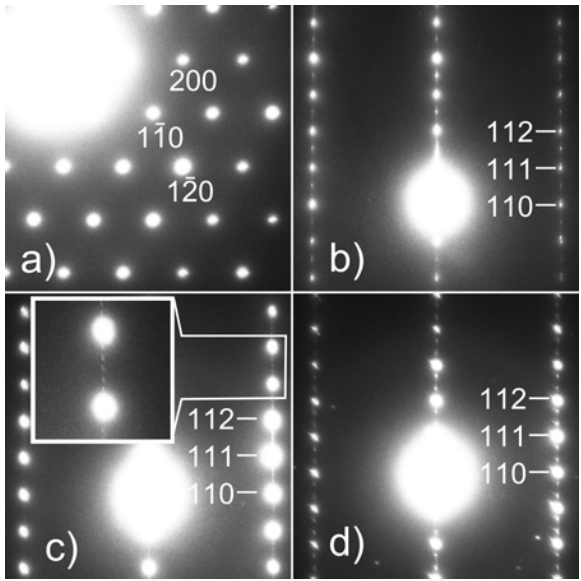

**Figure 3.**

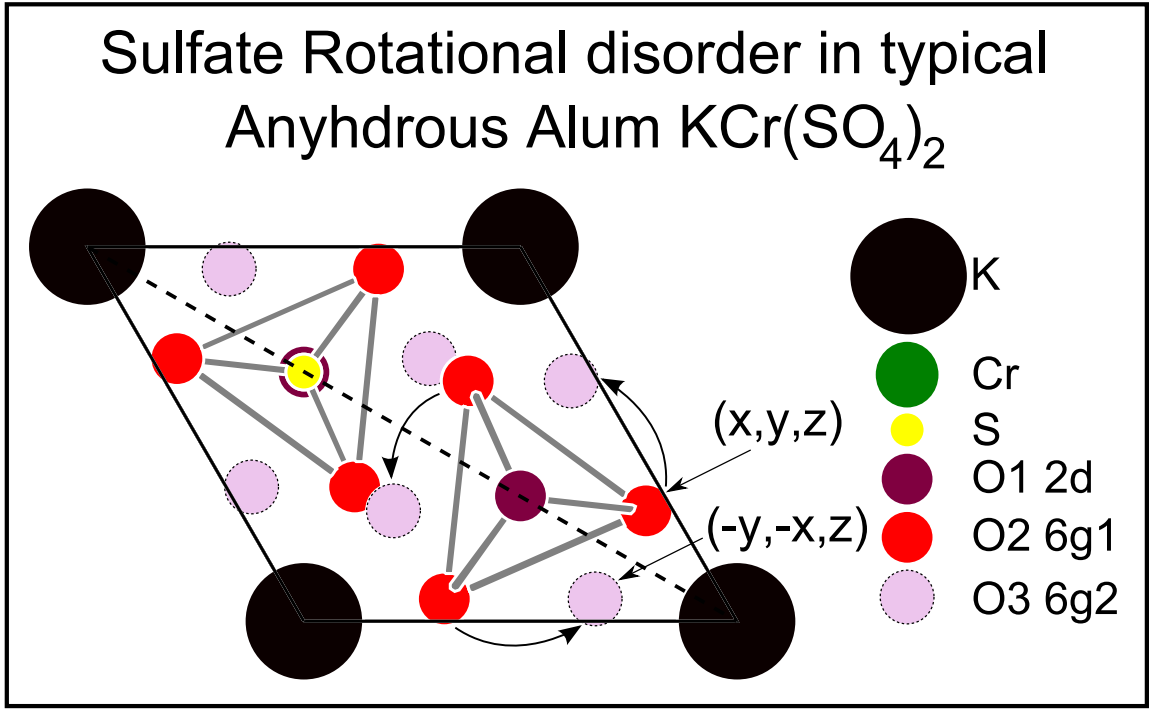

**Figure 4.**

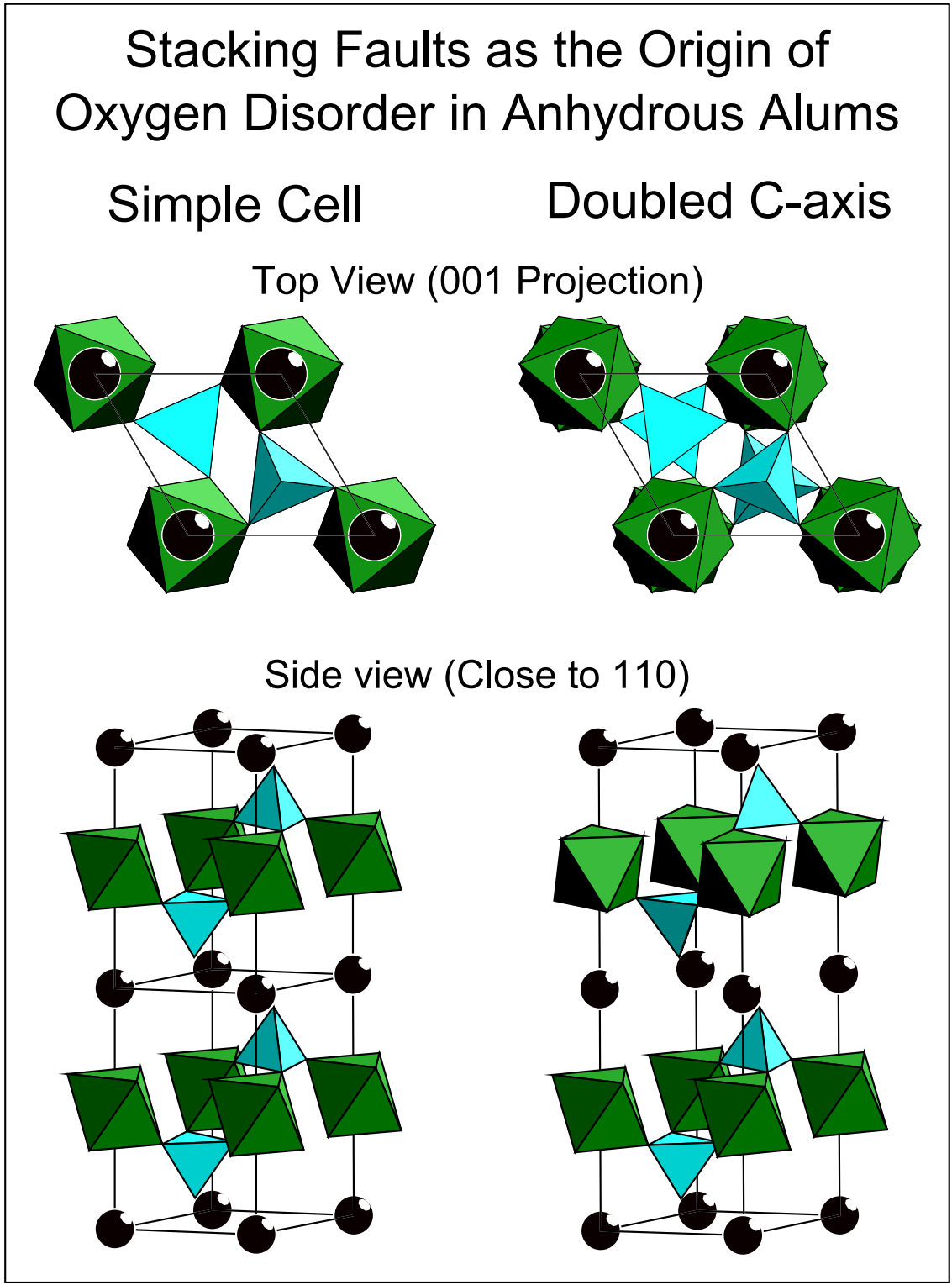

**Figure 5.**

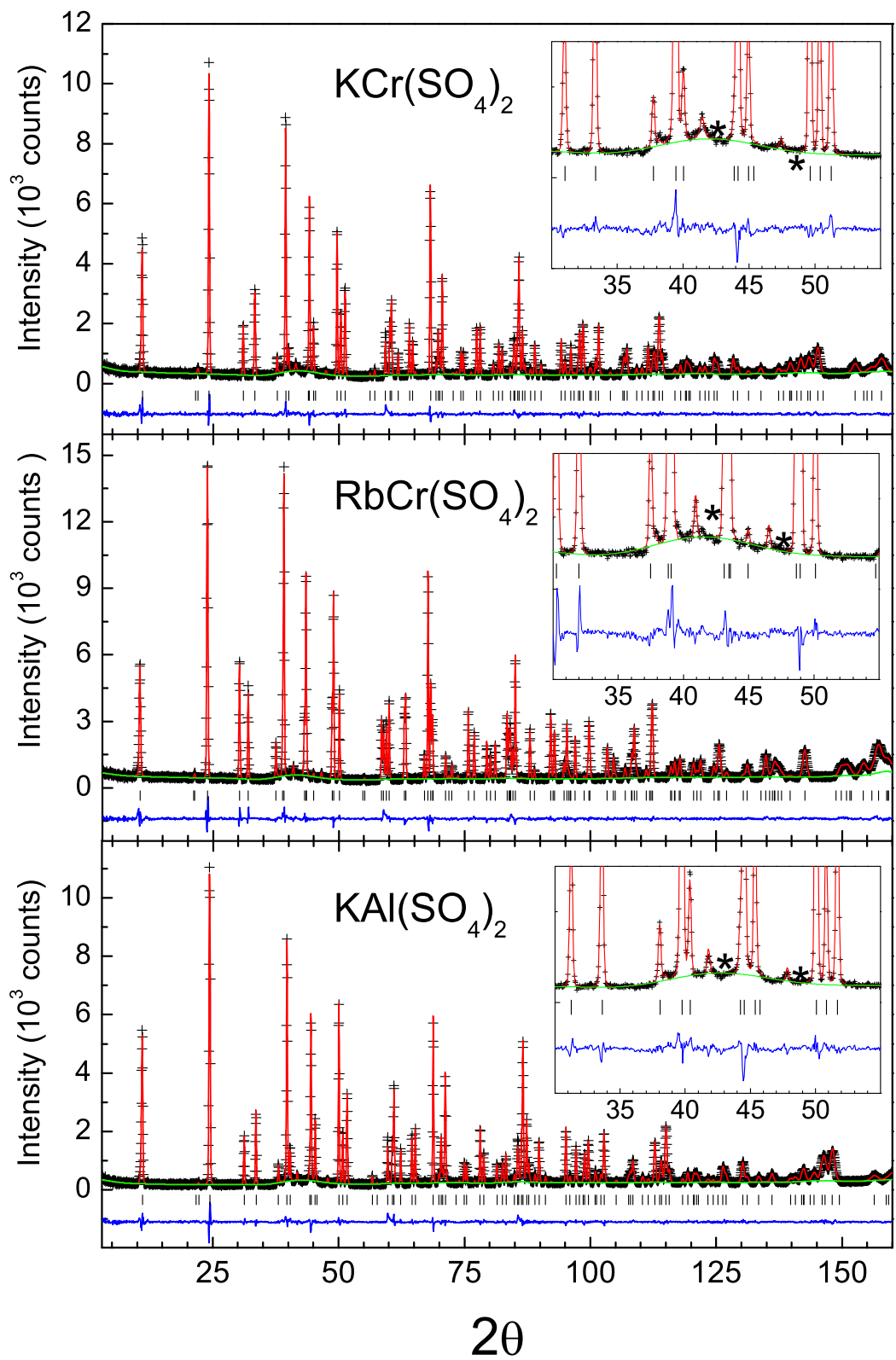

**Figure 6.**

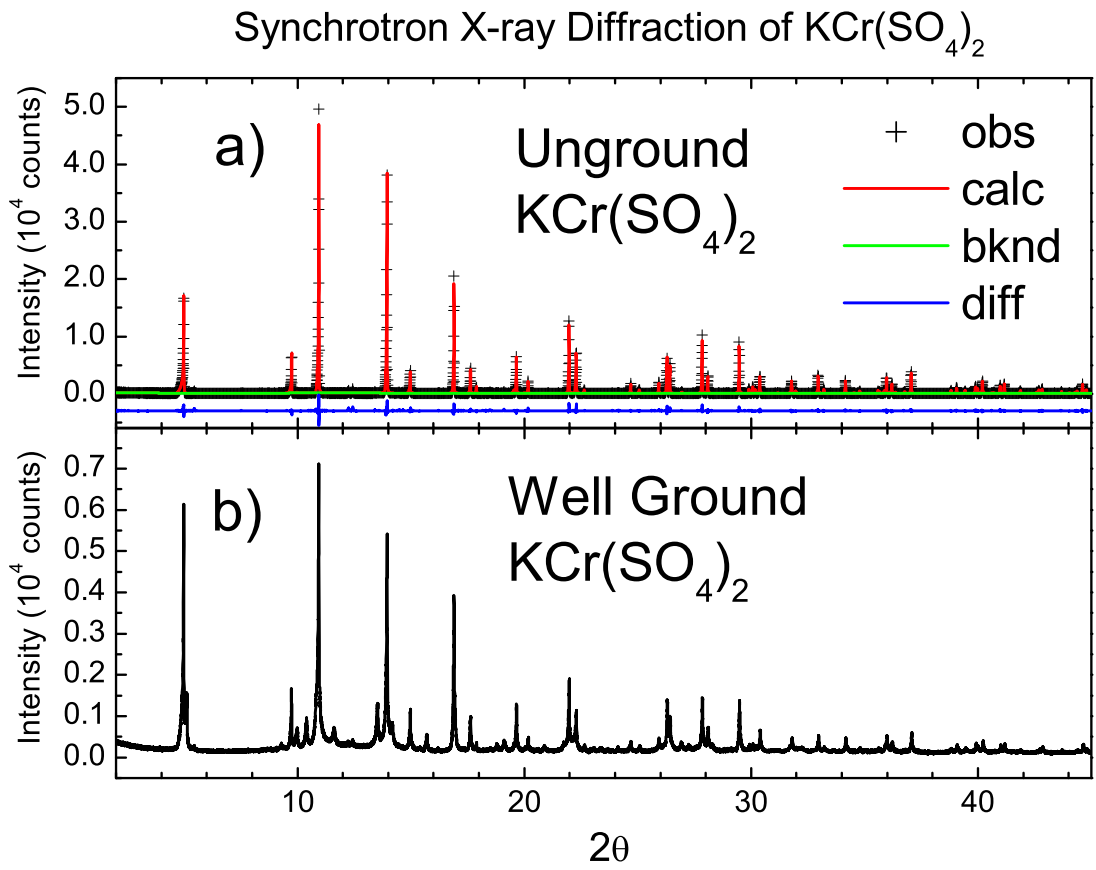

**Figure 7.**

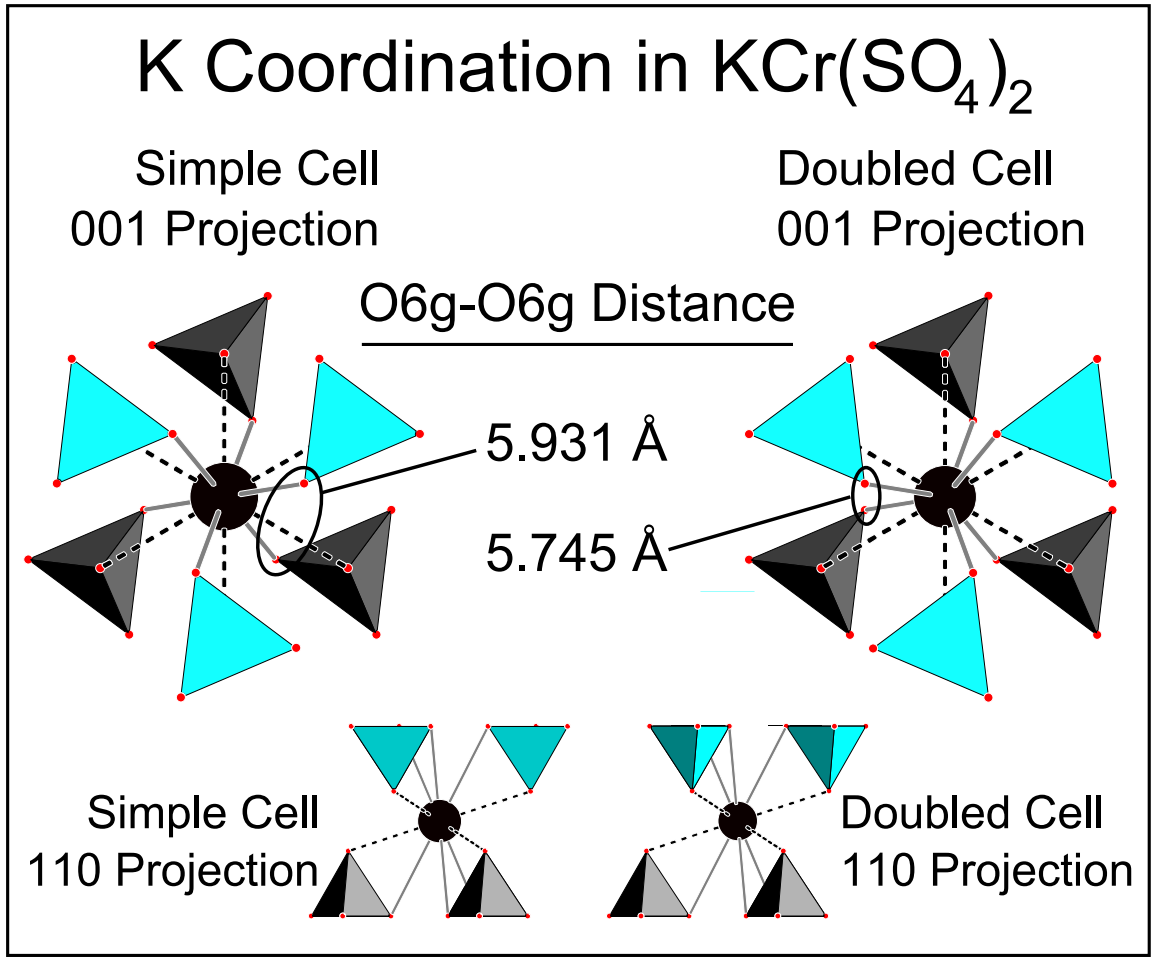

**Figure 8.**

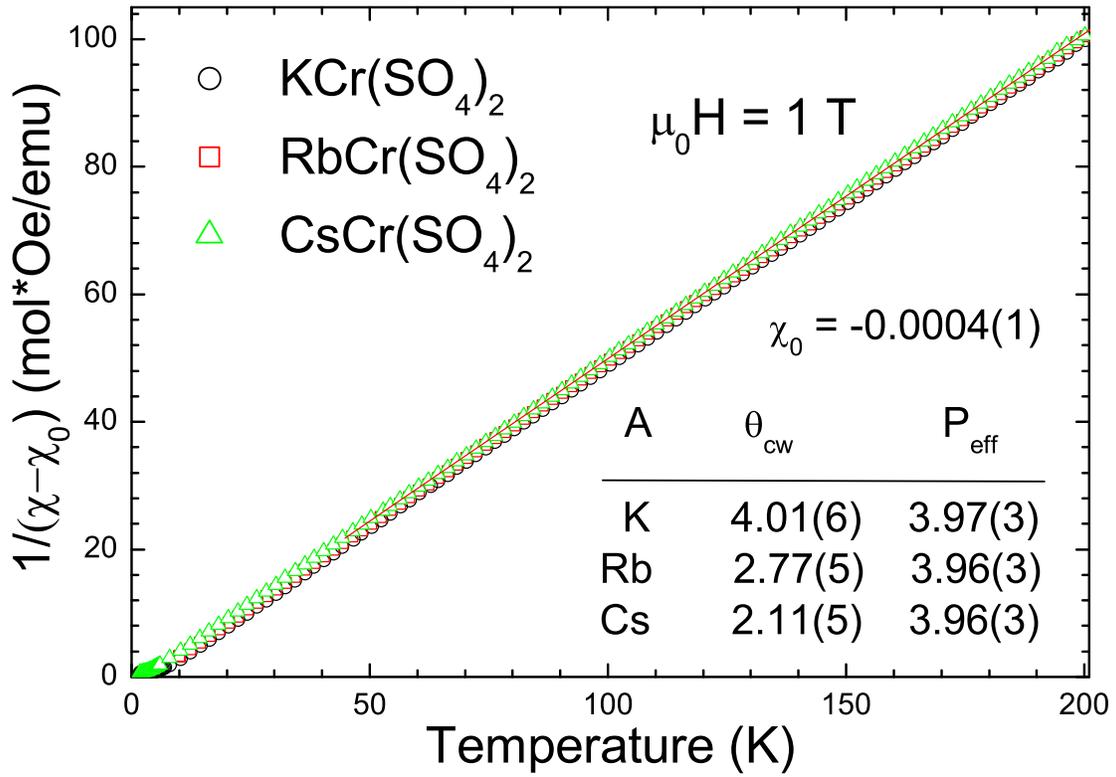

**Figure 9.**

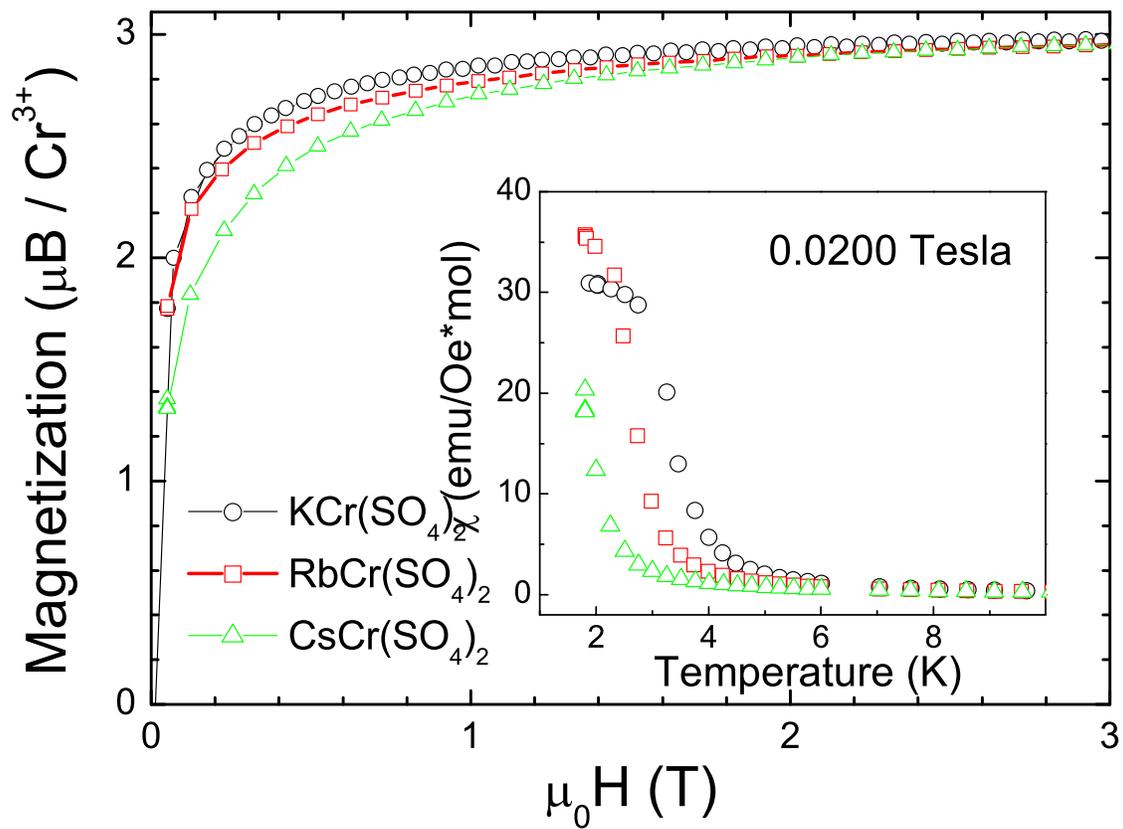

**Figure 10.**

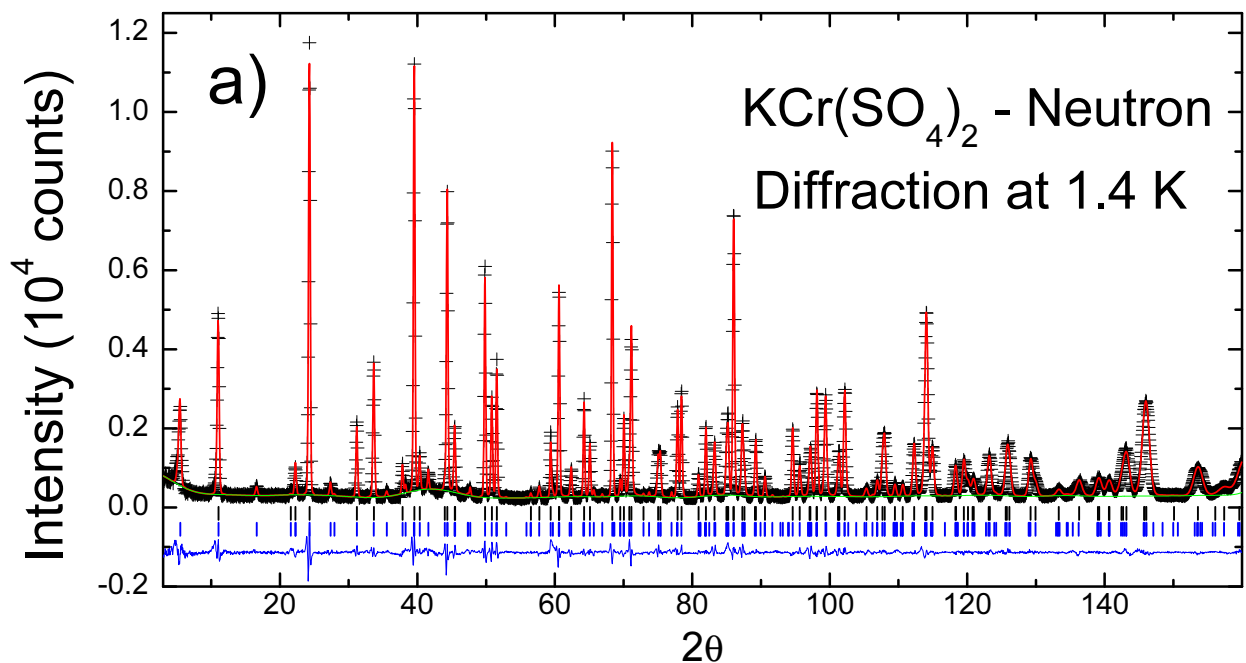

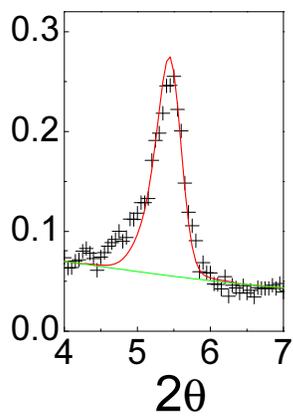

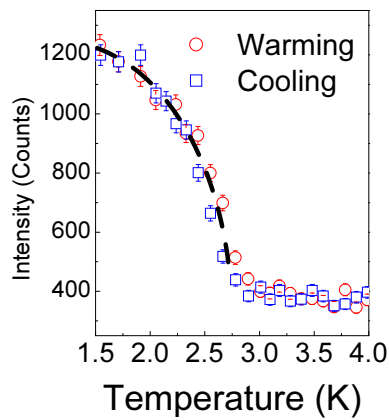

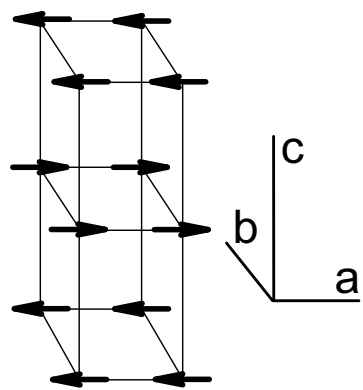